# Spatial period-doubling in Bose-Einstein condensates in an optical lattice

M. Machholm,[1,2,*] A. Nicolin,[3,4] C. J. Pethick[1], and H. Smith[2]

[1]*NORDITA, Blegdamsvej 17, DK-2100 Copenhagen Ø, Denmark*
[2]*Ørsted Laboratory, H. C. Ørsted Institute, Universitetsparken 5, DK-2100 Copenhagen Ø, Denmark.*
[3]*Niels Bohr Institute, Blegdamsvej 17, DK-2100 Copenhagen Ø, Denmark*
[4]*Politehnica, University of Bucharest, Romania.*
(Dated: November 14, 2003)

We demonstrate that there exist stationary states of Bose-Einstein condensates in an optical lattice that do not satisfy the usual Bloch periodicity condition. Using the discrete model appropriate to the tight-binding limit we determine energy bands for period-doubled states in a one-dimensional lattice. In a complementary approach we calculate the band structure from the Gross-Pitaevskii equation, considering both states of the usual Bloch form and states which have the Bloch form for a period equal to twice that of the optical lattice. We show that the onset of dynamical instability of states of the usual Bloch form coincides with the occurrence of period-doubled states with the same energy. The period-doubled states are shown to be related to periodic trains of solitons.

The properties of Bose-Einstein-condensed atoms in an optical lattice have been under intense investigation during the past few years. Some of the highlights include the observation of Josephson oscillations [1], Landau-Zener tunneling [2] and the quantum phase transition from a superfluid to a Mott insulator state [3]. On the theoretical side it has been found that the interaction between particles has dramatic effects on the band structure. When the interparticle interaction becomes sufficiently strong, "swallow tails" appear in the band structure of a one-dimensional lattice, both at the boundary of the Brillouin zone and at the zone center [4–7].

So far, theoretical considerations have focused on solutions to the Gross-Pitaevskii equation obeying the usual Bloch condition. For a one-dimensional (1D) lattice potential this implies that the condensate wave function $\psi$ has the form $\psi(x) = \exp(ikx)f(x)$, where $f(x)$ is a periodic function with the same period, $d$, as that of the optical lattice, $f(x+d) = f(x)$. In this paper we show that there exist stationary states with periods equal to a multiple of the lattice period, and that dynamical instability of the usual Bloch states sets in when there is a period-doubled state of the same energy and wave number.

First, we consider the case when the potential wells of the optical lattice are sufficiently deep that only a single orbital at each site need be considered. We shall further assume that the total number of particles on each site is sufficiently large and the tunneling strength sufficiently small that a mean-field Gross-Pitaevskii approach may be applied in which the complex amplitude of the condensate wave function on one site is influenced by those on neighboring sites. This discrete model was used in Ref. [8] to predict the breakdown of phase coherence in a chain of weakly coupled condensates. Within this model the condensate wave function $\psi$ is approximated by a superposition of wave functions $\Phi_j$ localized around each lattice site $j$, $\psi(\mathbf{r},t) = \sum_j \psi_j(t)\Phi_j(\mathbf{r})$. The coefficients $\psi_j$ depend on the site, while the wave functions $\Phi_j(\mathbf{r}) = \Phi(\mathbf{r} - \mathbf{r}_j)$ are assumed to be identical in form and normalized according to the condition $\int d\mathbf{r}\, |\Phi_j|^2 = 1$. The normalization condition is $\int d\mathbf{r}\, |\psi|^2 = N$, where $N$ is the total number of particles. If in evaluating this condition one neglects the overlap between wave functions localized at different sites, one finds $N = \sum_j N_j$, where $N_j = |\psi_j|^2$ is the number of particles on site $j$.

The Hamiltonian for the discrete model is [8]

$$H = -\sum_j K(\psi_j^* \psi_{j+1} + \psi_j \psi_{j+1}^*) + \sum_j \frac{1}{2}U|\psi_j|^4, \quad (1)$$

where the first term in (1) describes tunneling between neighboring sites, $K$ being the hopping parameter. The second term is the on-site particle interaction, the quantity $U = U_0 \int d\mathbf{r}|\Phi_j(\mathbf{r})|^4$ being the interaction energy between two atoms on the same site. $U_0$ is the effective interaction between two atoms, given in terms of the scattering length $a$ by $U_0 = 4\pi\hbar^2 a/m$, where $m$ is the atomic mass.

We consider stationary states with a fixed total number of particles. These are obtained by requiring that the variation of $H - \mu N$ with respect to $\psi_j^*$ vanish, where $\mu$ is the chemical potential. This yields

$$U|\psi_j|^2 \psi_j - K(\psi_{j+1} + \psi_{j-1}) - \mu \psi_j = 0. \quad (2)$$

In the following we focus on states in which the particle density is periodic with a period $2^p$ times the lattice spacing, with $p = 0, 1, 2, \ldots$. To calculate the energy per particle and the chemical potential we divide the 1D-lattice into cells, each containing $2^p$ sites. The number of particles within a cell is denoted by $N_c$, and $\nu = N_c/2^p$ is thus the average particle number per site.

We separate from $\psi_j$ a part which is a plane wave evaluated at the lattice point $jd$ where $d$ is the lattice period,

*[*]machholm@nordita.dk

$\psi_j = e^{ikjd}g_j$. Equation (2) then becomes

$$U|g_j|^2 g_j - K e^{ikd} g_{j+1} - K e^{-ikd} g_{j-1} - \mu g_j = 0. \quad (3)$$

Within the unit cell the complex amplitudes $g_j = |g_j|e^{i\phi_j}$ are thus related by the equations (3) with $j = 1, 2, \ldots, 2^p$. The periodicity is imposed by the requirement $g_{j+2^p} = g_j$. The number of particles within the unit cell is given by

$$N_c = \sum_{j=1}^{2^p} |g_j|^2. \quad (4)$$

Solutions for $p = 0$ and $1$ can be obtained analytically, while higher period-doubled states ($p = 2$ and $3$) are determined numerically.

For $p = 0$ the equation (3) with the boundary conditions $g_0 = g_1$ and $g_2 = g_1$ yields $U|g_1|^2 = \mu + 2K\cos kd$. Since $N_c = \nu = |g_1|^2$ the chemical potential is

$$\mu = -2K\cos kd + U\nu, \quad (5)$$

while the energy per particle, $\varepsilon$, obtained from (1) is

$$\varepsilon = -2K\cos kd + \frac{1}{2}U\nu. \quad (6)$$

The first term is the usual tight-binding expression for the energy of a Bloch state for a single particle.

For $p = 1$ the two equations (3) together with (4) are solved for $g_1$ and $g_2$ with the boundary conditions $g_0 = g_2$ and $g_3 = g_1$. Subtracting the two equations (3) we obtain

$$U(|g_1|^2 - |g_2|^2) = 2K(|g_2||g_1|^{-1}e^{i(\phi_2-\phi_1)} - |g_1||g_2|^{-1}e^{i(\phi_1-\phi_2)})\cos kd. \quad (7)$$

These equations are similar to those for self-trapped states of a condensate in a potential with two wells [9]. There is one class of solutions with $|g_1|^2 \neq |g_2|^2$. Since the left hand side of (7) is real, the phase difference $\phi_1 - \phi_2$ must be either $0$ or $\pi$. These solutions exist when $|\cos kd| \leq U\nu/2K$. This condition can always be satisfied, no matter how small the strength of the repulsive interaction $U$, provided $|k|$ is sufficiently close to $\pi/2d$. For $U\nu/2K > 1$ the period-doubled states extend over all of the Brillouin zone. Using the fact that, according to (4), $N_c = |g_1|^2 + |g_2|^2$ and solving for the magnitudes $|g_1|$ and $|g_2|$ we obtain the energy per particle as

$$\varepsilon = 2\frac{K^2}{U\nu}\cos^2 kd + U\nu. \quad (8)$$

The chemical potential $\mu = \partial(\nu\varepsilon)/\partial\nu$ obtained from (8) is given by $\mu = 2U\nu$, which is independent of $k$. This independence of $k$ is a special feature for $p = 1$ only, but the variations in $\mu$ for the states with longer periods are found to be very small.

There is, however, another class of solutions, since for $|k| = \pi/2d$, $|g_1| = |g_2|$ is a solution for arbitrary $\phi_1 - \phi_2$.

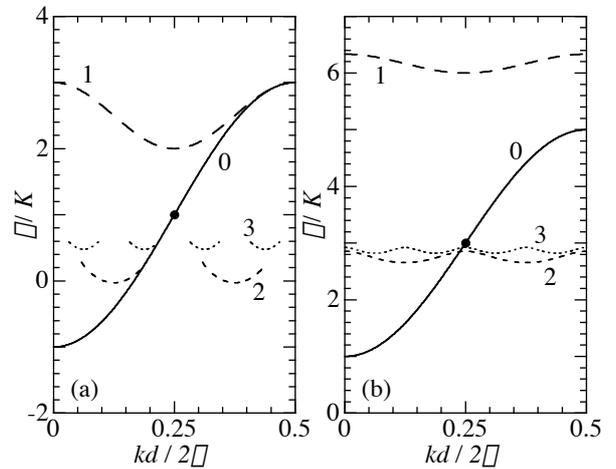

FIG. 1: Energy per particle in units of $K$ as a function of wave number $k$ for $p = 0, 1, 2$ and $3$ (indicated by labels). a) $U\nu/K = 2$. b) $U\nu/K = 6$. The $p = 1$ phase states are indicated by the solid circle at $k = \pi/2d$.

The energy per particle is $\epsilon = U\nu/2$, which is the same as that of the usual Bloch state for that wave number. These states may be constructed from the usual Bloch states at $|k| = \pi/2d$ by shifting the relative phase of the two sublattices by an arbitrary amount, and we shall refer to them as phase states. Observe that, while the particle number is the same on all sites, $g_j$ does not satisfy the usual Bloch condition, except for the particular case when $\phi_1 - \phi_2$ is a multiple of $2\pi$.

According to (8) the quantity $\varepsilon/K$ is a function only of the dimensionless parameter $U\nu/K$. We exhibit the energy bands for $U\nu/K = 2$ in Fig. 1a, and for $U\nu/K = 6$ in Fig. 1b. For simplicity we do not show the repetitions of the $p = 0$ band with different displacements in $k$ space that would appear if one were to represent the band in the reduced zone scheme corresponding to a cell size $2^p d$. Note that for $U\nu/K = 6$ (Fig. 1b) the $p = 1, 2$ and 3 bands are much narrower than the $p = 0$ band. The mean energy of the $p = 1$ band is roughly twice that of the others, reflecting the fact that for the $p = 1$ band the occupancy of every other site is nearly zero when $U\nu \gg K$.

The magnitude of the parameters $U$ and $K$ of the tight-binding model may be estimated in terms of the parameters of the optical lattice and the magnetic trap under realistic experimental conditions, as explained in Refs. [10] and [11]. It is convenient to express the energies in terms of the energy $E_R = \hbar^2\pi^2/2md^2$. If the lattice is created by oppositely directed laser beams, $E_R$ is the recoil energy, since then $d = \lambda/2$, where $\lambda$ is the wavelength of the lasers producing the optical lattice. For barrier heights between $10\,E_R$ and $20\,E_R$ one finds that the dimensionless parameter $U\nu/K$ ranges between 1 and 100 for typical densities used experimentally, since the effective mass, which is inversely proportional to $K$, increases strongly with increasing barrier height.

In order to investigate the stability of the solutions we expand the energy functional (1) to second order in the deviation $\delta\psi$ from the equilibrium solution $\psi^0$, following the methods used in Refs. [12], [8] and [7]. The solution is energetically stable if the second-order term is positive for all $\delta\psi$. We insert $\psi = \psi^0 + \delta\psi$ into (1) with a deviation of the general form $\delta\psi_j = e^{ikjd}(u_j e^{iqjd} + v_j^* e^{-iqjd})$. The first order terms vanish since $\psi^0$ satisfies (2).

We first consider the stability of the usual Bloch states ($p = 0$). In this case $u_j$ and $v_j$ are independent of the site index $j$. By carrying out the sum over $j$ we obtain a quadratic form in $u$ and $v$. The condition for energetic stability is that all eigenvalues of the matrix $B$ given by

$$B = \begin{pmatrix} U\nu + \Delta\varepsilon_+ & U\nu \\ U\nu & U\nu + \Delta\varepsilon_- \end{pmatrix} \quad (9)$$

are positive. Here $\Delta\varepsilon_\pm = \varepsilon(k \pm q) - \varepsilon(k)$ where $\varepsilon(k)$ is given by (6). This yields the energetic stability condition $\sin kd \tan kd < U\nu/2K$.

The condition for dynamical stability of the $p = 0$ states is that the eigenvalues of the matrix $\sigma_z B$ are real. This further yields the stability criterion $\cos kd > 0$, which was first derived in [13] and compared with measurements of superfluid current disruption in [14]. An important observation is that the onset of dynamical instability of the usual Bloch states occurs at $|k| = \pi/2d$ when the state becomes degenerate with the period-doubled phase states.

We turn now to period-doubled states with $p = 1$. Their stability may be investigated by the methods described above, but $B$ is now a $4 \times 4$ matrix. The matrix is Hermitian and the elements on the diagonal are of the form $U(|g_1|^2 - |g_2|^2)$ and $-U(|g_1|^2 - |g_2|^2)$, one of which is negative. Thus the matrix has negative eigenvalues, and the states are energetically unstable. However, there exist regions where the period-doubled states are dynamically stable if $U\nu/K \geq 2$, and these occur for $\pi/4d \leq |k| \leq 3\pi/4d$.

Our analysis of the period-doubling phenomena has so far been based on the discrete model, obtained from a tight-binding approximation to the Gross-Pitaevskii equation. Now we consider a continuum model by starting from the full Gross-Pitaevskii energy functional in the presence of an external potential $V(x)$, given by

$$V(x) = 2V_0 \cos^2(\pi x/d) = V_0 \cos(2\pi x/d) + V_0. \quad (10)$$

Apart from a constant term $nV_0$ the functional for the average energy density, $E$, is

$$E = \frac{1}{2d}\int_{-d}^{d} dx \left[ \frac{\hbar^2}{2m}\left|\frac{d\psi}{dx}\right|^2 + V_0 \cos\left(\frac{2\pi x}{d}\right)|\psi|^2 + \frac{1}{2}U_0|\psi|^4 \right], \quad (11)$$

and the average particle density $n$ is given by

$$n = \frac{1}{2d}\int_{-d}^{d} dx |\psi|^2. \quad (12)$$

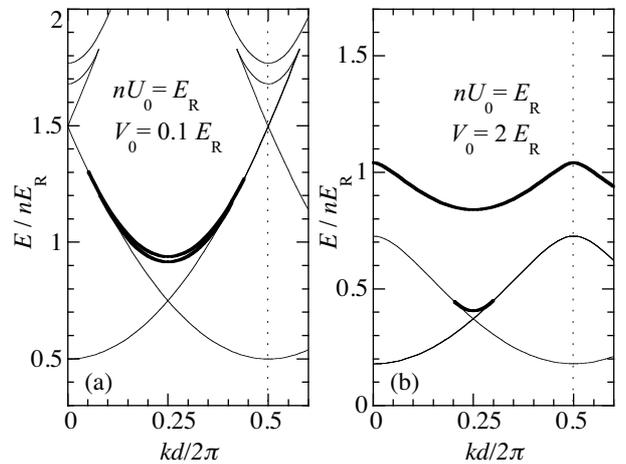

FIG. 2: Energy per particle in units of $E_R$ as a function of wave number $k$ for the lowest bands as obtained from the wave function (13). The bold curves correspond to spatial period-doubled states, and the thin curves to the usual Bloch states (See text).

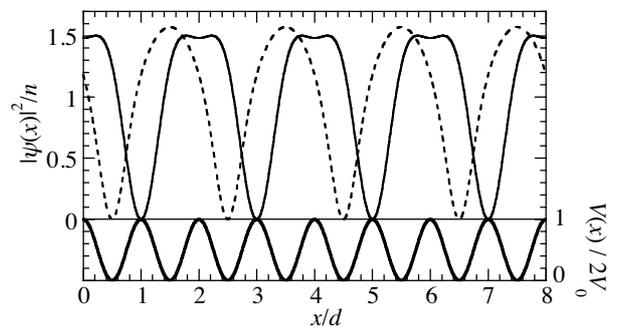

FIG. 3: Particle density $|\psi(x)|^2/n$ as a function of position for the lower- (full line) and higher-energy period-doubled states (dashed line) of Fig. 2a at $|k| = \pi/2d$. Lower panel: The periodic potential, Eq. (10).

To find solutions of the 1D Gross-Pitaevskii equation with periods $d$ and $2d$ we expand the condensate wave function $\psi(x)$ in plane waves by writing $\psi = e^{ikx}\sqrt{n}f(x)$, with

$$f(x) = \sum_{\ell=-\ell_{\max}}^{\ell_{\max}} a_\ell e^{i2\pi \ell x/d}, \quad (13)$$

where $\ell = 0, \pm 1/2, \pm 1, \pm 3/2, \ldots$. The coefficients $a_\ell$ satisfy the normalization condition $\sum_\ell |a_\ell|^2 = 1$. The stationary states of the system are obtained by demanding that the energy (11) be stationary with respect to variations in $\psi$, subject to the condition that the total number of particles remains constant, as in Ref. [7].

In Fig. 2 we show the band structure of the period-doubled states (thick lines), which are centered about $k = \pi/2d$. For comparison, we also show the band structure for the usual Bloch bands (thin lines). In a reduced zone scheme with lattice spacing $2d$ there are two such states for each wave number $k$ corresponding to energies $E(k)$ and $E(k+\pi/d)$ in the usual representation corresponding to a lattice spacing $d$.

A new feature compared to the discrete model is that the period-doubled phase states of the discrete model have become a band. For small $V_0$ (Fig. 2a) the two period-doubled bands are nearly degenerate and both merge continuously with the usual Bloch band. In the limit $V_0 \to 0$ the period-doubled bands are degenerate, and span a non-vanishing range in $k$-space which increases with increasing $nU_0$. Figure 2b shows bands for a stronger lattice potential. In this case only the lower band merges with the usual Bloch band, and the upper period-doubled band is lifted above the usual Bloch band as was the case in the discrete model. In order to compare the discrete and continuum models, we may estimate an effective value of $K$ by identifying either the curvatures of the bands at $k = 0$ or the band widths, and one finds that Figs. 2b and 1b correspond to similar physical conditions.

The period-doubled states may be understood in terms of trains of solitons just as in the case of states at the upper edge of swallow tails [7]. In the case of period-doubled states, there is one soliton for every two lattice spacings. To illustrate this, we show density distributions for period-doubled states at $k = \pi/2d$ in Fig. 3. For the upper band, dark solitons are centered on every second lattice site, as we also found in the discrete model, where $|g_j|^2 = N_c$ and $|g_{j+1}|^2 = 0$ at $k = \pi/2d$. For the lower band the dark solitons are centered at every second potential maximum. In the discrete model the lower band reduces to period-doubled phase states at $|k| = \pi/2d$.

We now consider the dynamical stability of the usual Bloch states. The wave number at which a condensate becomes dynamically unstable when accelerated slowly from rest has been calculated in Refs. [7] and [12] starting from the Gross-Pitaevskii equation (see in particular Fig. 7 of Ref. [7]). We have carried out numerical calculations for $nU_0$ and $V_0$ in the range 0-8 $E_R$, and find that dynamical instability sets in when the Bloch state becomes degenerate with the lowest period-doubled state, just as it did in the discrete model. However, because the lower band of period-doubled states has a non-zero range in $k$ in the continuum model, the value of $|k|$ for instability is larger than $\pi/2d$.

In conclusion, we have demonstrated the existence of a novel class of states which exhibit period-doubling in an optical lattice. On the basis of a discrete model appropriate to the tight-binding limit we have identified states with periods up to 8 times the period of the lattice, and solutions with other periods (such as 3, 5, etc. times the lattice period) may also occur. More generally, we have shown how the period-doubled states arise within the framework of the Gross-Pitaevskii equation. Solutions of the Gross-Pitaevskii equation with period $4d$ have been found by D. Diakonov [15].

Experimentally the period-doubled states could be detected by accelerating the condensate to a $k$-value slightly greater than the limit of dynamical instability and keeping it at fixed $k$ for some time before allowing the cloud to expand. Growth of a period-doubled component would be signaled by the appearance of extra peaks in the interference pattern lying midway between those expected for a lattice period $d$.

We thank Dmitri Diakonov, Mogens H. Jensen, Lars M. Jensen and Qian Niu for useful discussions and Magnus Johansen for correspondence. C.J.P. is grateful to the Aspen Center for Physics for hospitality. M.M. was supported by the Carlsberg Foundation.